\documentclass{aa}

\usepackage{graphicx}
\usepackage{yfonts}
\usepackage{xcolor}
\usepackage{txfonts}
\usepackage{subfigure}
\usepackage{float}
\usepackage{natbib,twoopt}
\usepackage{xspace}
\usepackage{balance}
\usepackage{upgreek}
\usepackage{multirow}
\usepackage[normalem]{ulem}
\usepackage{soul}
\usepackage[colorlinks=true, citecolor=blue, linkcolor=blue]{hyperref}
\usepackage[normalem]{ulem}

\usepackage[flushleft]{threeparttable}

\definecolor{redUniBo}{RGB}{187, 46, 41}

\def\apjl{ApJ Lett.}
\def\apjs{ApJ Suppl.}

\def\aap{A\&A}

\newcommand{\bp}{b_{\rm SZ}}

\bibpunct{(}{)}{;}{a}{}{,} 
\DeclareGraphicsExtensions{.eps,.ps,.pdf}

\defcitealias{Planck18}{Planck18}
\defcitealias{Planck15}{Planck15}

\begin{document}
\title{Mass bias and cosmological constraints from Planck cluster clustering}

\author
{G. F. Lesci\inst{\ref{1},\ref{2}}\orcid{0000-0002-4607-2830}
\and A. Veropalumbo\inst{\ref{3}}\orcid{0000-0003-2387-1194}
\and M. Sereno\inst{\ref{2},\ref{4}}\orcid{0000-0003-0302-0325}
\and F. Marulli\inst{\ref{1},\ref{2},\ref{4}}\orcid{0000-0002-8850-0303}
\and L. Moscardini\inst{\ref{1},\ref{2},\ref{4}}\orcid{0000-0002-3473-6716}
\and C. Giocoli\inst{\ref{2},\ref{4}}\orcid{0000-0002-9590-7961}
}

\offprints{G. F. Lesci \\ \email{giorgio.lesci2@unibo.it}}

\institute{
  Dipartimento di Fisica e Astronomia ``Augusto Righi'' - Alma Mater Studiorum
  Universit\`{a} di Bologna, via Piero Gobetti 93/2, I-40129 Bologna,
  Italy\label{1}
  \and INAF - Osservatorio di Astrofisica e Scienza dello Spazio di
  Bologna, via Piero Gobetti 93/3, I-40129 Bologna, Italy\label{2}
  \and Universit\`{a} degli Studi di Milano, via G. Celoria 16, I-20133 Milan, Italy\label{3}
  \and INFN - Sezione di Bologna, viale Berti Pichat 6/2, I-40127
  Bologna, Italy\label{4}
}

\date{Received --; accepted --}

\newcommand\bias{0.62}
\newcommand\biash{0.38}
\newcommand\biasup{0.14}
\newcommand\biaslow{0.11}

\abstract
   {}
   {We analysed the 3D clustering of the Planck sample of Sunyaev–Zeldovich (SZ) selected galaxy clusters, focusing on the redshift-space two-point correlation function (2PCF). We compared our measurements to theoretical predictions of the standard $\Lambda$ cold dark matter ($\Lambda$CDM) cosmological model, deriving an estimate of the Planck mass bias, $\bp$, and cosmological parameters.}
   {We measured the 2PCF of the sample in the cluster-centric radial range $r\in[10,150]$ $h^{-1}$Mpc, considering 920 galaxy clusters with redshift $z\leq0.8$. A Markov chain Monte Carlo analysis has been performed to constrain $\bp$, assuming priors on cosmological parameters from Planck Cosmic Microwave Background (CMB) results. We also adopted priors on $\bp$ from external data sets to constrain the cosmological parameters $\Omega_{\rm m}$ and $\sigma_8$.}
   {We obtained $(1-\bp)=\bias^{+\biasup}_{-\biaslow}$, which is in agreement with the value required to reconcile primary CMB and cluster count observations. By adopting priors on $(1-\bp)$ from external data sets, we derived results on $\Omega_{\rm m}$ that are fully in agreement and competitive, in terms of uncertainties, with those derived from cluster counts. This confirms the importance of including clustering in cosmological studies, in order to fully exploit the information from galaxy cluster statistics. On the other hand, we found that $\sigma_8$ is not constrained.}
   {}


\authorrunning{G. F. Lesci et al.}

\titlerunning{Mass bias and cosmology from Planck cluster clustering}

\maketitle

\section{Introduction}
Galaxy clusters are excellent tracers of the large scale matter distribution of the Universe, probing its geometry and evolution through their abundance and clustering \citep{Sereno15_cosm,Veropalumbo16,Costanzi19,Marulli21,Moresco21,To21,Lesci22_counts,Fumagalli22}. In particular, the formation and evolution of galaxy clusters can be theoretically described with high accuracy through numerical simulations. This allows the theoretical calibration of the cluster halo mass and bias functions \citep{ST1999,Sheth01,Tinker08,Tinker10,Despali16,Castro22} and the description of the cluster dark matter profiles \citep{NFW,BMO}, providing the link between cluster local and statistical properties. In addition, cluster masses can be measured with high precision through weak gravitational lensing \citep{Sereno17_PSZ2,Bellagamba19,Stern19} and X-ray observations \citep{Arnaud10,PlanckXX2013,Sereno17}. Also cluster abundance and clustering are suitable probes for mass calibration if a cosmological model is assumed \citep{Murata19,Chiu20,Lesci22}.\\
\indent As cosmological parameters are inferred with high precision in current cluster statistical analyses, accurate cluster mass calibrations are of critical importance. In fact, a not complete assessment of systematic uncertainties affecting the derived masses may lead to significant biases in the cosmological constraints \citep{PlanckCluserCounts15,Abbott20}. Simulations show that X-ray masses are typically 10-15 percent underestimated due to the assumption of hydrostatic equilibrium, for which bulk motions and turbulence in the intra-cluster medium are neglected \citep{Nagai07,Meneghetti10,Rasia12,LeBrun14}. Also weak lensing mass estimates can be biased, due to the inaccuracy of density profile models \citep{OguriHamana11}, baryonic effects influencing the halo concentration \citep{Henson17,Shirasaki18,Beltz-Mohrmann21}, halo orientation \citep{Becker11,Dietrich14,Zhang22} and projections \citep{Simet17,Melchior17}. As the biases in the weak lensing mass estimates are theoretically better understood, weak lensing observations are exploited to calibrate the main bias affecting X-ray masses, called hydrostatic bias, $b_{\rm h}$ \citep{vonDerLinden14,Hoekstra15,PlanckCluserCounts15,Smith16,Sereno17}. In particular, the relation between the X-ray mass, $M_X$, and the true mass, $M_{\rm tr}$, is usually expressed as $M_X=(1-b_{\rm h})M_{\rm tr}$. \\
\indent In this paper we focused on the mass bias of the Sunyaev–Zeldovich (SZ) selected Planck clusters \citep{PlanckCluserCounts15,PlanckClusterCat15}, referred to as the Planck mass bias, $\bp$. In fact, Planck cluster masses are expected to be biased low as they are derived from a scaling relation based on X-ray observations of 20 relaxed clusters at $z<0.2$ \citep{Arnaud10,PlanckXX2013}. We obtained an estimate of $\bp$ which is independent of lensing observations, by exploiting the monopole of the 3D two-point correlation function (2PCF) of the galaxy clusters present in the sample provided by \citet{PlanckClusterCat15}. In particular, we assumed a standard $\Lambda$ cold dark matter ($\Lambda$CDM) cosmological model, adopting the Cosmic Microwave Background (CMB) constraints on cosmological parameters from \citet{Planck18} as priors. In addition, we adopted the same priors on $(1-\bp)$ used in the Planck cluster count analysis carried out by \citet{PlanckCluserCounts15}, in order to constrain the matter density parameter, $\Omega_{\rm m}$, and the amplitude of the matter power spectrum, $\sigma_8$. \\
\indent The statistical analyses presented in this paper are performed with the CosmoBolognaLib\footnote{\url{https://gitlab.com/federicomarulli/CosmoBolognaLib/}} \citep[CBL;][]{cbl}, a set of \textit{free software} C++/Python numerical libraries for cosmological calculations. Specifically, both the measurements and the statistical Bayesian analyses are performed with the CBL v$6.1$.

\indent The paper is organised as follows. In Section \ref{sec:data} we describe the data set and the methods used to estimate the 2PCF of the sample. In Section \ref{sec:modelling} we describe the 2PCF model, focusing on the dependence of the effective bias on the mass-observable scaling relation. In Section \ref{sec:results} we show our constraints on $\bp$ and we detail the cosmological analysis, while in Section \ref{sec:summary} we draw our conclusions.

\section{Data set and 2PCF measurement}\label{sec:data}

\begin{figure}[t!]
\centering
\includegraphics[width = \hsize, height = 7.3cm] {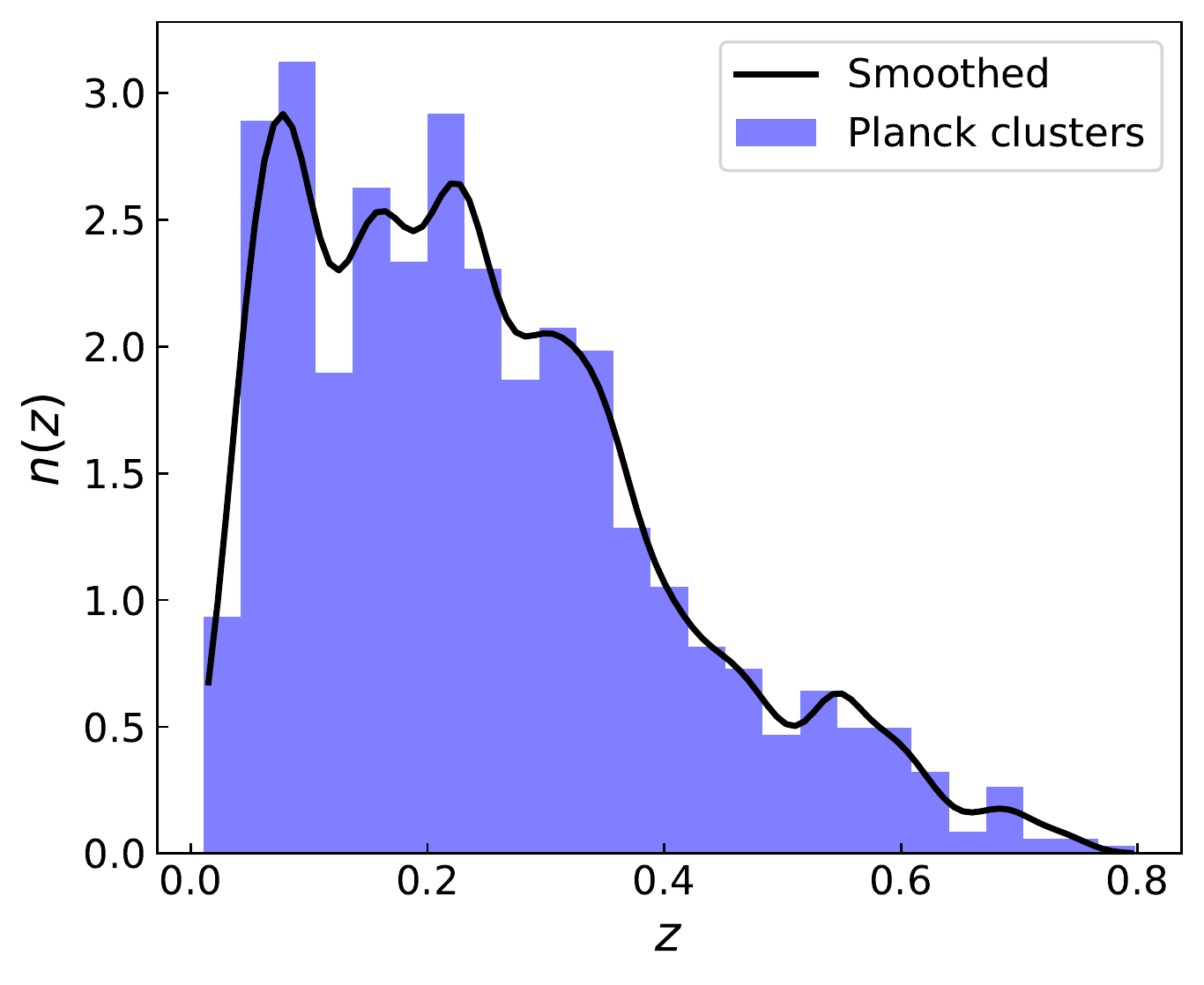}
\caption{Redshift distribution $n(z)$ of the galaxy clusters considered in the analysis. The blue histogram shows the observed binned $n(z)$, while the black curve represents $n(z)$ smoothed with a Gaussian kernel (with rms equal to 0.02), used to build up the random catalogue.}
\label{fig:nz}
\end{figure}

\subsection{The Planck cluster sample}
Following \citet{PlanckCluserCounts15}, we based our analysis on the cosmological sample consisting of detections by the MMF3 matched filter \citep{Melin06,MMF} derived from the general Planck full-mission Sunyaev–Zeldovich catalogue \citep[PSZ2,][]{PlanckClusterCat15}. We considered clusters having a confirmed counterpart in external data sets and an assigned redshift estimate \citep[see Table 9 in][]{PlanckClusterCat15}, with a redshift limit $z\leq0.8$, for a total of 920 clusters. We applied this redshift cut to exclude 5 clusters that are isolated with respect to the bulk of the redshift distribution, hindering the derivation of a reliable smoothed redshift distribution, which is necessary for the construction of the random sample (see Section \ref{sec:random}). In addition, differently from \citet{PlanckCluserCounts15}, we did not apply any cut in signal-to-noise ratio (S/N). This does not imply any potential problems due to the reliability of the selection function at low S/N, as our model does not rely on assumptions on the sample completeness (see Section \ref{sec:b_eff}). 

\begin{figure*}[t!]
\centering
\includegraphics[width = \hsize-1.5cm, height = 7.cm] {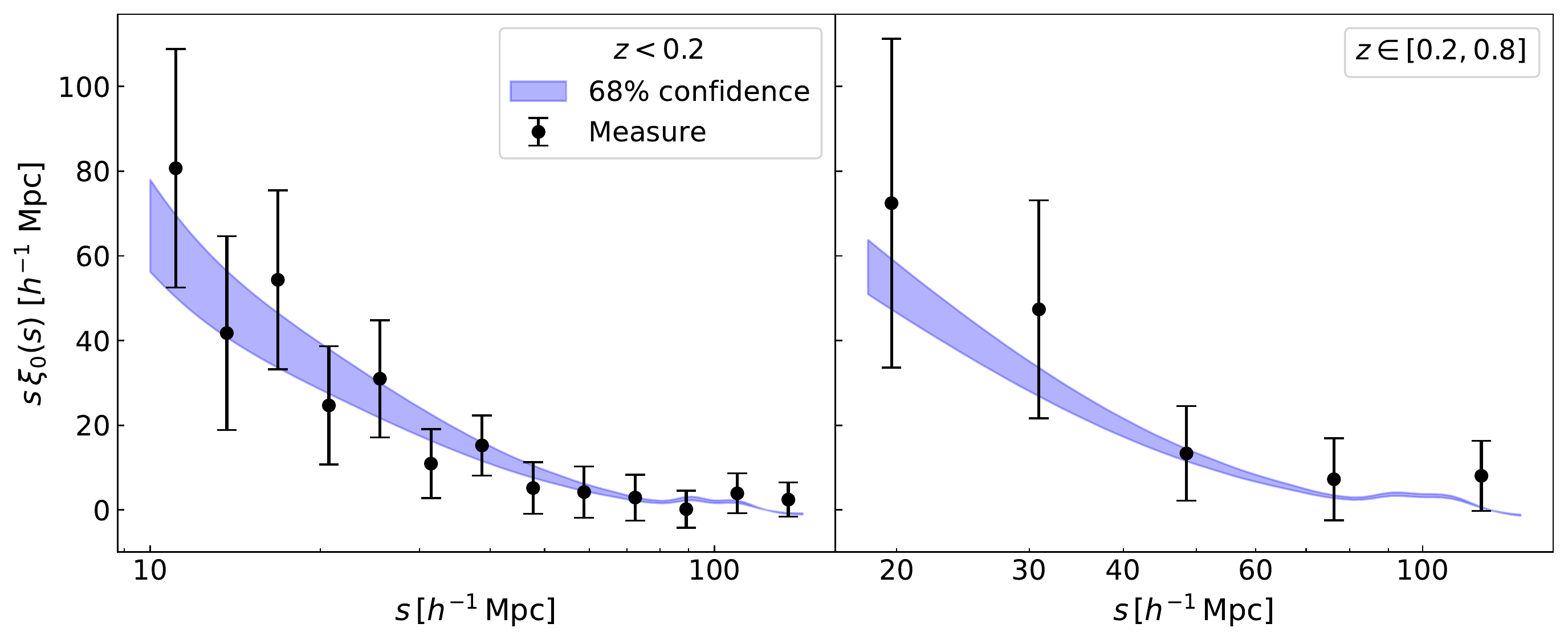}
\caption{Redshift-space 2PCF monopole (black dots) of the Planck clusters in the spatial range $s\in[10,150]$ $h^{-1}$Mpc for $z<0.2$ (left plot), and $s\in[15,150]$ $h^{-1}$Mpc for $z\in[0.2,0.8]$ (right plot). The blue bands represent the model 68\% confidence level derived from the posterior of the free parameters considered in the analysis described in Section \ref{sec:results:bh}.}
\label{fig:xi}
\end{figure*}

\subsection{Random catalogue}\label{sec:random}
The random catalogue used for the 2PCF measurement is 100 times larger than the Planck cluster sample. We smoothed the observed redshift distribution, $n(z)$, with a Gaussian kernel having rms equal to 0.02 (see Fig.\ \ref{fig:nz}). Then we extracted random redshifts from such distribution. Random R.A.-Dec pairs have been extracted by following the sample angular selection function. It consists of the combination of the MMF3 survey mask\footnote{\url{https://irsa.ipac.caltech.edu/data/Planck/release_2/ancillary-data/HFI_Products.html}}, namely $\mathcal{M}_{\rm s}$, the hole mask excluding contaminated regions (e.g.\ by stars, large galaxies, ...), $\mathcal{M}_{\rm h}$, and the error function completeness. Both $\mathcal{M}_{\rm s}$ and $\mathcal{M}_{\rm h}$ are equal to 0 if the region is masked, otherwise they are equal to 1. The error function completeness is defined as \citep{PlanckClusterCat13}
\begin{equation}\label{eq:erf_compl}
P(d|Y_{500},\sigma_{Yi}(\theta_{500}),q) = \frac{1}{2}\left[ 1+\text{erf}\left(\frac{Y_{500}-q\sigma_{Yi}(\theta_{500})}{\sqrt{2}\sigma_{Yi}(\theta_{500})}\right) \right],
\end{equation}
where $d$ is the Boolean detection state, $\text{erf}(x)$ is the Gauss error function, $Y_{500}$ and $\theta_{500}$ are the observed SZ signal and the detection angular
scale within a critical radius $R_{500}$, respectively, while $\sigma_{Yi}$ is the standard deviation of pixels for a given patch $i$, computed by following \citet{Melin06}, and $q$ is the S/N threshold. As we did not apply any S/N cut to the sample, $q$ corresponds to the minimum threshold adopted by \citet{PlanckClusterCat15} in the detection process, namely $q=4.5$. In Eq.\ \eqref{eq:erf_compl}, we assumed the sample mean values of $Y_{500}$ and $\theta_{500}$. We verified that using the median values of such quantities does not introduce significant variations in the final results. Then we extracted random angular positions, for each of which we sampled a number in the range $[0,1]$. In case such number was higher than the product of $\mathcal{M}_{\rm s}$, $\mathcal{M}_{\rm h}$ and $P(d|Y_{500},\sigma_{Yi},q)$, the random angular position was rejected. As an alternative to the error function completeness in Eq.\ \eqref{eq:erf_compl}, we weighted the pairs in the 2PCF estimator (described in Section \ref{sec:measure}) by $1/\sigma^{\rm norm}_{Yi}$, where $\sigma^{\rm norm}_{Yi}$ is equal to $\sigma_{Yi}$ divided by its minimum value, namely $\sigma^{\rm norm}_{Yi}=\sigma_{Yi}/\text{min}(\sigma_{Yi})$. We verified that this approach provides results that are fully in agreement with what derived from the application of the error function completeness.

\subsection{Clustering measurement}\label{sec:measure}
We estimated the redshift-space 2PCF monopole using the \citet[][LS]{LS1993}
estimator,
\begin{equation}\label{eq:LS}
\xi(s) = \frac{N_{RR}}{N_{DD}}\frac{DD(s)}{RR(s)} -
2\frac{N_{RR}}{N_{DR}}\frac{DR(s)}{RR(s)}+1\,,
\end{equation}
where $DD(s)$, $RR(s),$ and $DR(s)$ are the number of data-data, random-random, and data-random pairs with separation $s\pm\Delta s$, respectively, while $N_{DD}$, $N_{RR}$ , and $N_{DR}$ are the total number of data-data, random-random, and data-random pairs, respectively. To convert the observed coordinates into the comoving ones, we assumed the cosmological parameters by \citet{Planck18}, TT, TE, EE+lowE+lensing (referred to as \citetalias{Planck18} hereafter). The LS estimator is extensively used in clustering analyses as it is unbiased with minimum variance for an infinitely large random sample and when $|\xi|\ll1$ \citep{Hamilton1992, Kerscher2000, Labatie10, Keihanen19}.\\
\indent Specifically, we measured the 2PCF considering two redshift bins, namely $z<0.2$ and $z\in[0.2,0.8]$, containing 407 and 513 galaxy clusters, respectively. We considered the cluster-centric radial range $s\in[10,150]$ $h^{-1}$Mpc, excluding from the analysis the 2PCF measure at $s<15$ $h^{-1}$Mpc in the second redshift bin due to the lack of data-data pairs. We estimated the covariance matrix, including the cross-covariance between radial and redshift bins, through a bootstrap procedure. In particular, we considered 200 angular regions and two redshift regions, corresponding to the redshift bins, and resampled the observed and random catalogues 2000 times. We corrected the inverted covariance matrix following \citet{HSS}. In Fig.\ \ref{fig:xi} we show the measured 2PCF monopole, $\xi_0$. We did not include the other non-zero multipoles in the analysis, as we verified that their contribution is negligible.

\section{Modelling}\label{sec:modelling}
\begin{table*}[t]
\caption{\label{tab1}Free parameters considered in the analysis detailed in Section \ref{sec:results:bh}.}
  \centering
    \begin{tabular}{l c c r} 
      Parameter & Description & Prior & Posterior \\ 
      \hline
      \rule{0pt}{4ex}
      $\bp$ & Planck mass bias & [-2,\,0.9] & $\biash^{+\biasup}_{-\biaslow}$\\ \rule{0pt}{2.5ex}
      $\log Y_*$ & Normalisation of the mass-observable relation & $\mathcal{N}(-0.19,0.02)$ & --- \\ \rule{0pt}{2.5ex}
      $\alpha$ & Slope of the mass-observable relation & $\mathcal{N}(1.79,0.08)$ & --- \\ \rule{0pt}{2.5ex}
      $\beta$ & Redshift evolution of the mass-observable relation & $\mathcal{N}(0.66,0.50)$ & --- \\ \rule{0pt}{2.5ex}
      $\sigma_{\ln Y}$ & Intrinsic scatter of the mass-observable relation & $\mathcal{N}(0.173,0.023)$ & ---\\ 
    \end{tabular}
  \tablefoot{In the third column, the priors on the parameters are listed. In particular, a range between square brackets represents a uniform prior, while $\mathcal{N}(\mu,\sigma)$ stands for a Gaussian prior with mean $\mu$ and standard deviation $\sigma$. In the fourth column, we show the median values of the 1D marginalised posteriors, along with the 16th and 84th percentiles. The posterior distributions of $\log Y_*$, $\alpha$, $\beta$, and $\sigma_{\ln Y}$ are not shown since these parameters are not constrained in our analysis.}
\end{table*}

We modelled the 2PCF of Planck clusters by accounting for geometric and redshift-space distortions. In addition, different to what was done in the Planck cluster counts analysis by \citet{PlanckCluserCounts15}, our model does not rely on assumptions on the sample completeness. We show in Section \ref{sec:results} that this approach leads to constraints on $\bp$ and cosmological parameters that are fully in agreement with those derived by \citet{PlanckCluserCounts15} and \citet{Planck18}.

\subsection{Two-point correlation function model}
The $l$-th order 2PCF multipole, $\xi_l(s)$, can be expressed as follows,
\begin{equation}
\xi_l(s)=i^l\int_{-\infty}^\infty\frac{{\rm d}k}{2\pi^2}\,k^2P_l(k)j_l(ks),
\end{equation}
where $j_l$ is the spherical Bessel function of order $l$, and $P_l$ is the redshift-space matter power spectrum multipole of order $l$,
\begin{equation}\label{eq:Pl}
P_l(k)=\frac{2l+1}{2\alpha^2_\perp\alpha_\parallel}\int_{-1}^1{\rm d}\mu \, P(k',\mu')L_l(\mu).
\end{equation}
In Eq.\ \eqref{eq:Pl},  $L_l$ is the Legendre polynomial of order $l$, and $\mu$ is the line of sight cosine. Moreover, in Eq.\ \eqref{eq:Pl} we accounted for the \citet[][AP]{AP} geometric distortions, caused by the assumption of a fiducial cosmology used to convert the cluster observed coordinates into comoving ones in Eq.\ \eqref{eq:LS}. Specifically, $k'$ and $\mu'$ have the following functional forms \citep{Beutler14},
\begin{equation}
  k^\prime = \frac{k}{\alpha_\perp} \left[1 + \mu^2 \left(
    \frac{\alpha^{2}_\perp}{\alpha^{2}_\parallel} -1
    \right)\right]^{1/2} \, ,
\end{equation}

\begin{equation}
  \mu^\prime = \mu\,\frac{\alpha_\perp}{\alpha_\parallel} \left[1 +
    \mu^2 \left( \frac{\alpha^{2}_\perp}{\alpha^{2}_\parallel} - 1
    \right)\right]^{-1/2}\, ,
\end{equation}
where $\alpha_\perp$ and $\alpha_\parallel$ are expressed as
\begin{equation}
  \alpha_\parallel = \frac{H^{\rm fid}(z)r_{\rm s}^{\rm
      fid}(z_d)}{H(z)r_{\rm s}(z_d)} \, ,
\end{equation}

\begin{equation}
  \alpha_\perp = \frac{D_A(z)r_{\rm s}^{\rm fid}(z_d)}{D_A^{\rm
      fid}(z)r_{\rm s}(z_d)} \, .
\end{equation}
Here $H^{\rm fid}(z)$ and $D_A^{\rm fid}(z)$ are the fiducial values for the Hubble constant and angular diameter distance, respectively, and $r_{\rm s}^{\rm fid}(z_{\rm d})$ is the fiducial sound horizon at the drag redshift, $z_{\rm d}$. We stress that the AP correction takes place only in the cosmological analysis described in Section \ref{sec:results:cosm}: in fact for the derivation of $\bp$, detailed in Section \ref{sec:results:bh}, we fixed the cosmological parameters to the fiducial ones.
In Eq.\ \eqref{eq:Pl}, $P(k',\mu')$ is the redshift-space dark matter power spectrum expressed as \citep{Taruya10}:
\begin{multline}\label{eq:Pk}
  P(k', \mu') = D_{FoG}(k', \mu', f, \sigma_{\rm v})\,\biggl[b_{\rm eff}^2P_{\delta\delta}(k') + 2fb_{\rm eff}\mu'^2P_{\delta\theta}(k')+\\
  + f^2\mu'^4P_{\theta\theta}(k') + b_{\rm eff}^3A(k', \mu', f) + b_{\rm eff}^4B(k', \mu', f)\biggr] \, ,
\end{multline}
where $P_{\delta\delta}$, $P_{\theta\theta}$, and $P_{\delta\theta}$ are the real-space auto power spectra of density and velocity divergence, and their cross power spectrum, respectively. These spectra are estimated
in the Standard Perturbation Theory (SPT), consisting in expanding the statistics as a sum of infinite terms, corresponding to the $n$-loop corrections \citep[see e.g.][]{GilMarin12}. Considering corrections up to the first-loop order, the power spectrum can be modelled as follows:
\begin{equation}
P^{\rm SPT}(k) = P_{\rm L}(k) + P^{(1)}(k) = P_{\rm L}(k) + 2P_{13}(k) + P_{22}(k),
\end{equation}
where the leading order term, $P_{\rm L}(k)$, is the linear matter power spectrum, computed with \texttt{CAMB}\footnote{\url{https://camb.info/}} \citep{CAMB}, while the one-loop correction terms are computed with the \texttt{CPT Library}\footnote{\url{http://www2.yukawa.kyoto-u.ac.jp/~atsushi.taruya/cpt_pack.html}} \citep{CPT1}. In Eq.\ \eqref{eq:Pk}, $D_{FoG}(k', \mu', f, \sigma_{\rm v})$ is a Gaussian damping function representing the Fingers of God effect, having the following functional form:
\begin{equation}
D_{FoG}(k', \mu', f, \sigma_{\rm v}) = e^{-k'^2\mu'^2f^2\sigma_{\rm v}^2},
\end{equation}
where $f$ is the linear growth rate, and $\sigma^2_{\rm v}$ is the linear velocity dispersion, computed as \citep{Taruya10}:
\begin{equation}\label{eq:sigmav}
\sigma^2_{\rm v} = \frac{1}{3}\int\frac{{\rm d}^3\boldsymbol{k}}{(2\pi)^3}\,\frac{P_{\rm L}(k)}{k^2}.
\end{equation}
In Eq.s \eqref{eq:Pk} and \eqref{eq:sigmav}, $P_{\rm L}(k)$ is computed at the mean redshift of the cluster sub-sample in the given redshift bin. In addition, in Eq.\ \eqref{eq:Pk}, $b_{\rm eff}$ is the effective bias, defined in Section \ref{sec:b_eff}, while the functions $A(k', \mu', f)$ and $B(k', \mu', f)$ are correction terms derived from SPT \citep{Taruya10,deLaTorre12,Garcia20}. 

\subsection{Effective bias}\label{sec:b_eff}
The effective bias, $b_{\rm eff}$, has the following functional form,
\begin{align}\label{eq:bias_eff}
  b_{\rm eff} = \frac{1}{N_{\rm cl}}\sum_{j=1}^{N} b(Y^{\rm ob}_{500,j},z_j^{\rm ob}),
\end{align}
where $N_{\rm cl}$ is the number of clusters in the sample, $Y^{\rm ob}_{500,j}$ and $z_j^{\rm ob}$ are the observed SZ signal and redshift, respectively, of the $j$-th cluster, and $b(Y^{\rm ob}_{500,j},z^{\rm ob}_j)$ is expressed as
\begin{align}\label{eq:bias}
\nonumber
b(Y^{\rm ob}_{500,j},z^{\rm ob}_j) &= \frac{1}{n(Y^{\rm ob}_{500,j},z^{\rm ob}_j)} \\ \nonumber &\times \int_0^\infty {\rm d}M_{500}\,\, \frac{{\rm d}n(M_{500},z^{\rm ob}_j)}{{\rm d}M_{500}}\,\, b(M_{500},z^{\rm ob}_j) \\ &\times \int_0^\infty {\rm d}Y_{500}\, P(Y_{500}|M_{500},z^{\rm ob}_j)\, P(Y_{500}|Y^{\rm ob}_{500,j}) \,,
\end{align}
where $b(M_{500},z)$ is the halo bias, for which the model by \citet{Tinker10} is assumed, while $P(Y_{500}|Y^{\rm ob}_{500,j})$ is a Gaussian whose mean is $Y^{\rm ob}_{500,j}$ and its root mean square deviation (rms) is given by the error on $Y^{\rm ob}_{500,j}$. In addition, $P(Y_{500}|M_{500},z)$ is a log-normal whose mean is given by the mass-observable scaling relation and its rms is given by the intrinsic scatter, $\sigma_{\ln Y}$, 
\begin{equation}
P(\ln Y_{500}|M_{500},z) = \frac{1}{\sqrt{2\pi}\sigma_{\ln Y}} e^{-\ln^2(Y_{500}/\bar{Y}_{500})/(2\sigma^2_{\ln Y})}\,.
\end{equation}
Specifically, following \citet{PlanckCluserCounts15}, we assumed $\sigma_{\ln Y}$ to be independent of $Y_{500}$ and redshift, and the expected value of SZ signal, $\bar{Y}_{500}$, can be expressed as
\begin{equation}\label{eq:scalrel}
E^{-\beta}(z)\left[\frac{D_{\rm A}^2(z) \bar{Y}_{500}}{\mathrm{10^{-4}\,Mpc^2}}\right] =  Y_* \left[ {h \over 0.7}
  \right]^{-2+\alpha} \left[\frac{(1-\bp)\,
    M_{500}}{6\times10^{14}\,{\rm M}_\odot}\right]^{\alpha},
\end{equation}
where $E(z) \equiv H(z)/H_0$, with $H(z)$ being the Hubble function and $H_0$ the Hubble constant, $D_A(z)$ is the angular diameter distance, $h\equiv H_0/100$, $\bp$ is the Planck mass bias, while $Y_*$, $\alpha$ and $\beta$ are the scaling relation parameters. In addition $n(Y^{\rm ob}_{500,j},z^{\rm ob}_j)$ in Eq.\ \eqref{eq:bias} is expressed as
\begin{align}
n(Y^{\rm ob}_{500,j},z^{\rm ob}_j) &= \int_0^\infty{\rm d}M_{500} \,\, \frac{{\rm d}n(M_{500},z^{\rm ob}_j)}{{\rm d}M_{500}} \nonumber\\ &\times \int_0^\infty {\rm d}Y_{500}\, P(Y_{500}|M_{500},z^{\rm ob}_j)\, P(Y_{500}|Y^{\rm ob}_{500,j}) \,,
\end{align}
where ${\rm d}n(M_{500},z)/{\rm d}M_{500}$ is the halo mass function, for which the model by \citet{Tinker08} is assumed. 

\subsection{Likelihood}\label{sec:like}
For the Bayesian analysis performed in this work, a standard Gaussian likelihood was considered,
\begin{equation}
\mathcal{L} \propto \exp(-\chi^2/2)\,,
\end{equation}
with
\begin{align}
&\chi^2=\sum_{i=1}^N\sum_{j=1}^N
  \left(\xi_i^d-\xi_i^m \right)\, C_{i,\,j}^{-1}\, \left(
  \xi_j^d-\xi_j^m\right)\,,
\end{align}
where $N$ is the number of comoving separation bins in which the 2PCF is computed, $d$ and $m$ indicate {data} and {model}, respectively, and $C_{i,\,j}^{-1}$ is the inverse of the {covariance matrix}. As detailed in Section \ref{sec:measure}, $C_{i,\,j}$ is derived through a bootstrap resampling.

\section{Results}\label{sec:results}
Based on the methods outlined in Sections \ref{sec:data} and \ref{sec:modelling}, we carried out an analysis of the redshift-space 2PCF monopole of the Planck cluster sample \citep{PlanckClusterCat15}. Specifically, in Section \ref{sec:results:bh} we detail the derivation of the $(1-\bp)$ constraint, performed by assuming the \citetalias{Planck18} cosmological results as priors. In Section \ref{sec:results:cosm} we present the constraints on cosmological parameters, obtained by assuming priors on $\bp$ from external data sets.

\subsection{Constraint on $\bp$}\label{sec:results:bh}
\begin{figure}[t!]
\centering
\includegraphics[width = \hsize-0.5cm, height = 13.4cm] {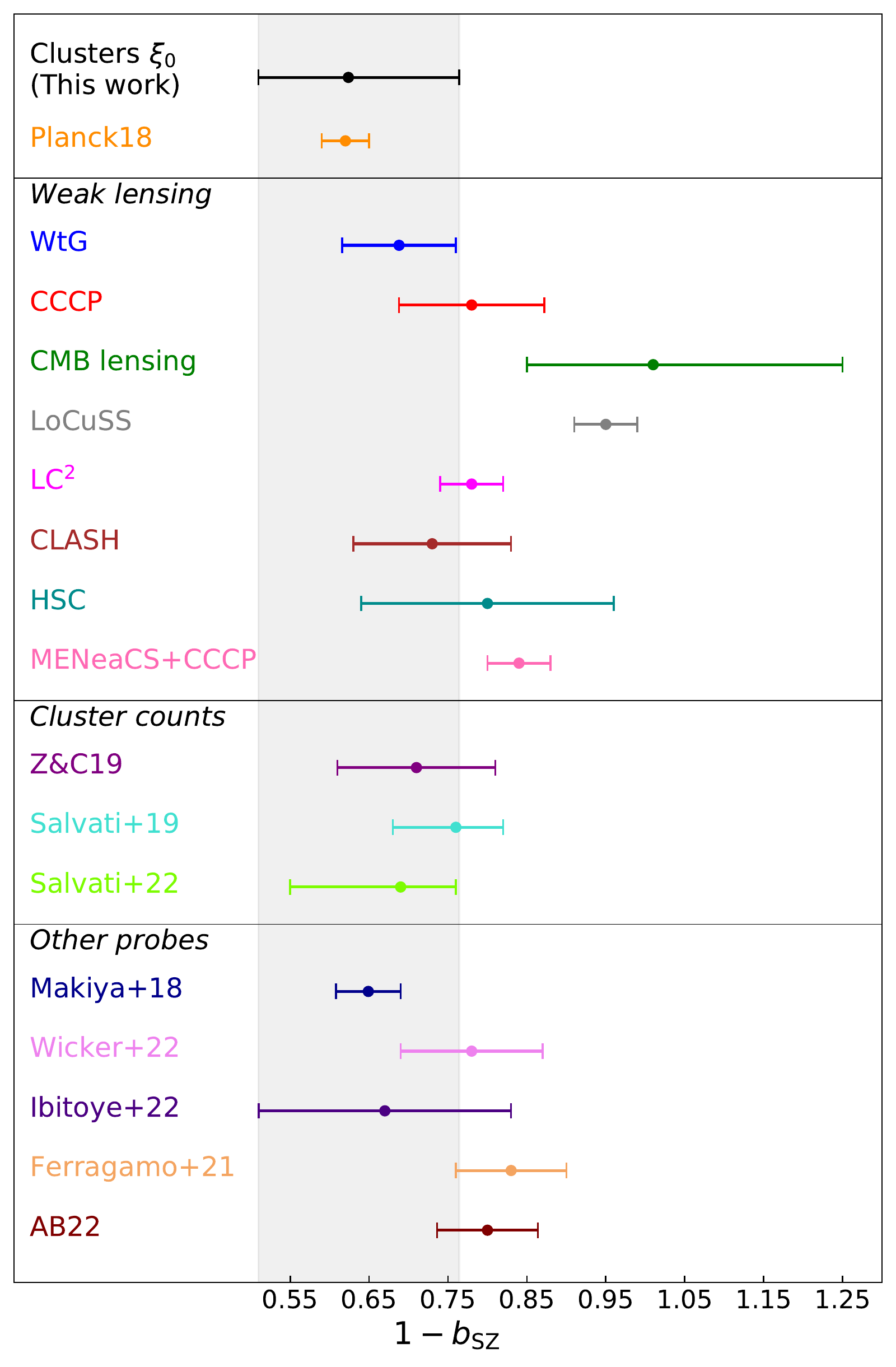}
\caption{Comparison of the results on $(1-\bp)$ with literature. Median, 16th and 84th percentiles are shown. The black dot shows the constraint derived in this work. Then, in order from top to bottom, the following results are shown: \citet{Planck18} (orange), \citet{vonDerLinden14} (blue) \citet{Hoekstra15} (red), \citet{PlanckCluserCounts15} (dark green), \citet{Smith16} (grey), \citet{Sereno17} (magenta), \citet{Penna-Lima17} (brown), \citet{Medezinski18} (cyan), \citet{Herbonnet20} (pink), \citet{Zubeldia19} (purple), \citet{Salvati19} (turquoise), \citet{Salvati22} (green), \citet{Makiya18} (dark blue), \citet{Wicker22} (violet), \citet{Ibitoye22} (indigo), \citet{Ferragamo21} (orange), \citet{Aguado22} (dark brown).}
\label{fig:bh}
\end{figure}

In order to derive a constraint on the Planck mass bias, $\bp$, we fixed the cosmological parameters to the \citetalias{Planck18} median values. We also assumed the priors on the mass-observable scaling relation parameters in Eq.\ \eqref{eq:scalrel}, namely $Y_*$, $\alpha$, $\beta$, and $\sigma_{\ln Y}$, adopted by \citet{PlanckCluserCounts15}. In particular, this scaling relation was derived from X-ray observations of 20 relaxed clusters at $z<0.2$ \citep{Arnaud10,PlanckXX2013}. Finally, we assumed a large flat prior on $\bp$. In Table \ref{tab1} we summarise the priors used for this analysis, along with the result on the mass bias, namely $(1-\bp)=\bias^{+\biasup}_{-\biaslow}$. The corresponding effective bias estimates are $b_{\rm eff}=4.61^{+0.39}_{-0.36}$ and $b_{\rm eff}=6.46^{+0.35}_{-0.37}$ for $z<0.2$ and $z\in[0.2,0.8]$, respectively. The constraint on $(1-\bp)$ is lower compared to what predicted by numerical simulations \citep{Nagai07,Piffaretti08,Meneghetti10,Rasia12,LeBrun17,Henson17,Gianfagna22}, but in line with what found by \citet{Planck18}. We remark that our constraint is dominated by the 2PCF signal measured at low redshift, as we obtained $(1-\bp)=0.67^{+0.22}_{-0.13}$ for $z<0.2$ and $(1-\bp)=0.58^{+0.55}_{-0.31}$ for $z\in[0.2,0.8]$. \\
\indent In Fig.\ \ref{fig:bh} we show a comparison between our constraint on $(1-\bp)$ and the results obtained from literature. In presence of systematic uncertainties, we added them in quadrature to the statistical ones. By combining primary CMB likelihood and cluster counts, \citet{Planck18} derived $(1-\bp)=0.62\pm0.03$ (orange dot in Fig.\ \ref{fig:bh}), which is fully in agreement with our result. Regarding the Planck mass estimates derived from galaxy weak lensing, we found a $1\sigma$ agreement with Weighting the Giants \citep[WtG;][]{vonDerLinden14}, Canadian Cluster Comparison Project \citep[CCCP;][]{Hoekstra15}, Literature Catalogs of weak Lensing Clusters of galaxies \citep[LC$^2$;][]{Sereno17}, Cluster Lensing And Supernova survey with Hubble \citep[CLASH;][]{Penna-Lima17}, Subaru Hyper Suprime-Cam \citep[HSC;][]{Medezinski18}. We found instead only a $2\sigma$ agreement with the results from the Local Cluster Substructure Survey \citep[LoCuSS;][]{Smith16}, Multi Epoch Nearby Cluster Survey (MENeaCS) combined
with updated mass weak lensing estimates in CCCP \citep[MENeaCS+CCCP;][]{Herbonnet20}, and with the result obtained from CMB lensing by \citet{PlanckCluserCounts15}. When comparing our results to other analyses based on cluster counts, we found a $1\sigma$ agreement with \citet{Zubeldia19}, \citet{Salvati19}, and \citet{Salvati22}. Concerning the results derived from the power spectra of the Planck thermal Sunyaev–Zeldovich effect, our constraint is in agreement within $1\sigma$ with \citet{Makiya18} and \citet{Ibitoye22}. We also found a good agreement with the constraint by \citet{Wicker22}, based on measurements of the cluster gas mass fraction. Regarding the hydrostatic bias estimates from dynamical masses, we found a $1\sigma$ agreement with \citet{Ferragamo21} and \citet{Aguado22}. In Section \ref{sec:results:sys} we discuss the impact of the adopted modelling choices on our result, finding that the derived constraint on $\bp$ is robust with respect to the investigated systematic uncertainties. \\
\indent As many observational studies claimed the presence of a redshift dependence of the hydrostatic bias \citep{Smith16,Sereno17,Salvati19,Salvati22,Wicker22}, we investigated this possibility by expressing $\bp$ as follows:
\begin{equation}
\bp = \eta\,\left( \frac{1+z}{1+z_{\rm piv}} \right)^\zeta,
\end{equation}
where $z_{\rm piv}=0.25$ is the mean redshift of the sample, $\eta$ is the normalisation, and $\zeta$ parametrises the redshift dependence of the mass bias. It turns out that our analysis does not constrain $\zeta$, implying that it is not necessary to explain our data. We stress that the redshift dependence of $\bp$ was derived from cluster statistics only in the case of a strong prior on the total value of $\bp$, with a significant dependence on the sample \citep{Salvati19,Salvati22,Wicker22}.

\subsubsection{Assessment of systematics}\label{sec:results:sys}
To assess the robustness of the constraint on $\bp$ derived in Section \ref{sec:results:bh}, we included the power spectrum damping due to redshift uncertainties in the analysis. As redshift errors are not quoted in Planck data products, we expressed this damping by means of a free parameter. Specifically, we replaced Eq.\ \eqref{eq:sigmav} by the following expression
\begin{equation}
\sigma_{\rm v,tot} = \sqrt{\sigma_{\rm v}^2+\sigma_{\text{v},\,z}^2}\,,
\end{equation}
where $\sigma_{\rm v}$ is defined in Eq.\ \eqref{eq:sigmav}, while $\sigma_{\text{v},\,z}$ is the velocity dispersion caused by redshift errors, having the following functional form
\begin{equation}
\sigma_{\text{v},\,z} \equiv \frac{c\,\sigma_z(1+\bar{z})}{H(\bar{z})}\,.
\end{equation}
In this equation, $\bar{z}$ is the mean redshift of the sub-sample in a given redshift bin, $c$ is the speed of light, $H(z)$ is the Hubble function, while $\sigma_z$ is the typical redshift uncertainty of the sample. By assuming a flat prior on $\sigma_z$, namely $\sigma_z\in[0,0.1]$, we derived $\sigma_z=0.003^{+0.002}_{-0.002}$, in line with the fact that most of the cluster redshifts are spectroscopic, and $(1-\bp)=0.65^{+0.15}_{-0.12}$, which is fully in agreement with our previous result. \\
\indent In addition, we analysed the 2PCF monopole of the Planck union catalogue, containing the clusters detected with the three detection algorithms adopted by \citet{PlanckClusterCat15}. By assuming the same sample selections and bins of redshift and radius described in Section \ref{sec:data}, we found $(1-\bp)=0.59^{+0.12}_{-0.09}$, which is in line with the constraint derived in Section \ref{sec:results:bh}. This implies the independence of our result on the adopted cluster detection algorithm. We also performed the analysis by considering the clusters in the MMF3 sample with $S/N>6$, and for which the \texttt{COSMO} entry in the union catalogue is set to \texttt{’T’}, following \citet{PlanckCluserCounts15}. Due to the poorer statistics in this case, we analysed the 2PCF in a single bin of redshift including clusters with $z\leq0.8$, for a total of 430 objects. As the modelling provides reduced $\chi^2$ estimates that are not close to 1, we conclude that in this case the 2PCF signal does not allow a reliable constraint on $\bp$. \\
\indent In order to further assess the robustness of our results on $(1-\bp)$, we computed the 2PCF model at the sample median redshifts for each redshift bin, instead of adopting the mean redshift as discussed in Section \ref{sec:modelling}. In this way we derived a shift of the median $(1-\bp)$ of $\sim0.006\sigma$. In addition, the reduction of the 2PCF radial range to $s\in[15,150]$ $h^{-1}$Mpc or to $s\in[10,90]$ $h^{-1}$Mpc implies comparable results, namely shifts of the median $(1-\bp)$ lower than $\sim0.6\sigma$, and variations of the $1\sigma$ interval extension lower than $\sim50$\%. We also checked the impact of a change in the definition of the effective bias, $b_{\rm eff}$, assuming the median of the halo bias distribution instead of considering its mean, as done in Eq.\ \eqref{eq:bias_eff}. In this case we obtained a shift of the median $(1-\bp)$ corresponding to $\sim0.5\sigma$. As the tests described above showed shifts of the median $\bp$ that are within $1\sigma$ of the constraint presented in Section \ref{sec:results:bh}, we can conclude that our results are robust with respect to the investigated modelling choices.

\subsection{Constraints on cosmological parameters}\label{sec:results:cosm}
\begin{figure}[t!]
\centering
\includegraphics[width = \hsize-1.cm, height = 7.cm] {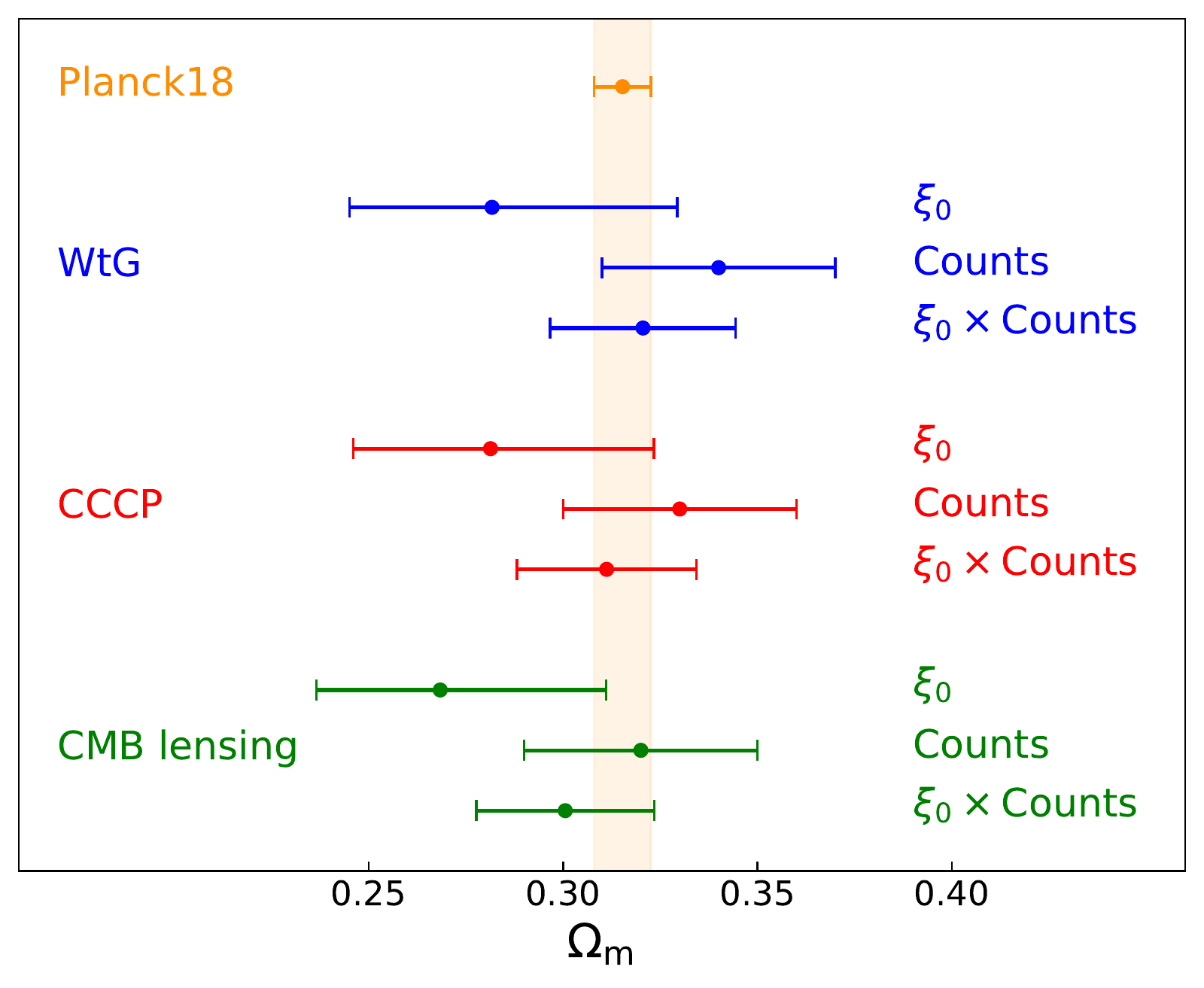}
\caption{Comparison of the results on $\Omega_{\rm m}$ obtained by assuming flat priors on $\Omega_{\rm m}$ and $\sigma_8$, along with external priors on $\bp$, namely WtG \citep[blue,][]{vonDerLinden14}, CCCP \citep[red,][]{Hoekstra15} and CMB lensing \citep[green,][]{PlanckCluserCounts15}. For each $\bp$ prior, the result on top was derived from the cluster clustering measurements presented in this paper, the middle one refers to the cluster counts analysis by \citet{PlanckCluserCounts15}, while the bottom one represents the combination of Planck cluster clustering and counts. The constraint from \citetalias{Planck18} is shown in orange.}
\label{fig:OmegaM}
\end{figure}

To further investigate the consistency of our modelling choices with those adopted by \citet{PlanckCluserCounts15}, we performed a cosmological analysis aiming at constraining $\sigma_8$ and $\Omega_{\rm m}$ simultaneously, by assuming the same priors on $\bp$ considered by \citet{PlanckCluserCounts15}. Specifically, we assumed large flat priors for $\sigma_8$ and $\Omega_{\rm m}$, while for the other cosmological parameters we assumed the same values from \citetalias{Planck18} used in the previous section. It turns out that $\sigma_8$ is not constrained through this analysis, while we found $\Omega_{\rm m}=0.28^{+0.05}_{-0.04}$ with the WtG $\bp$ prior, $\Omega_{\rm m}=0.28^{+0.04}_{-0.03}$ with the CCCP prior, and $\Omega_{\rm m}=0.27^{+0.04}_{-0.03}$ with the CMB lensing prior (see Fig.\ \ref{fig:OmegaM}). These results are fully consistent and competitive, in terms of uncertainties, with those derived by \citet{PlanckCluserCounts15}. We also derived an estimate of $\Omega_{\rm m}$ from the combination of cluster clustering and counts, by assuming them to be statistically independent: with respect to the analysis based on counts only, the uncertainty on $\Omega_{\rm m}$ is reduced by a factor of $\sim25\%$-$30\%$.
This confirms the importance of including cluster clustering in cosmological analyses \citep[see also][]{Sartoris16,Fumagalli22}, in order to fully exploit the cluster statistics information. In addition, we note that significant changes in the value of $\bp$ do not imply significant variations in the $\Omega_{\rm m}$ posteriors, similar to what found by \citet{PlanckCluserCounts15}.

\section{Summary and discussion}\label{sec:summary}
In this work we analysed the 3D 2PCF monopole of the galaxy clusters detected by \citet{PlanckClusterCat15}, focusing on the estimate of the Planck mass bias, $\bp$. Following \citet{PlanckCluserCounts15}, we based our analysis on the cosmological sample consisting of detections by the MMF3 matched filter \citep{Melin06,MMF}, considering clusters with a confirmed counterpart in external data sets and having an assigned redshift estimate, with a redshift limit $z\leq0.8$, for a total of 920 clusters. Differently from \citet{PlanckCluserCounts15}, we did not apply any cut in S/N to the sample. This does not imply any potential problems due to the reliability of the selection function at low S/N, as our model does not rely on assumptions on the sample completeness. \\
\indent By analysing the 2PCF in the redshift bins $z<0.2$ and $z\in[0.2,0.8]$, within the cluster-centric radial range $r\in[10,150]$ $h^{-1}$Mpc, we derived $(1-\bp)=\bias^{+\biasup}_{-\biaslow}$. This result is fully in agreement with what found by \citet{Planck18} by combining primary CMB likelihood and Planck cluster counts. Thus we confirmed that Planck cluster statistics provides values of $\bp$ that are lower compared to what predicted by numerical simulations \citep{Nagai07,Piffaretti08,Meneghetti10,Rasia12,LeBrun17,Henson17,Gianfagna22}. As redshift errors are not quoted in Planck data products, we also included the power spectrum damping due to redshift uncertainties by means of a free parameter representing the typical redshift error, namely $\sigma_z$. Thus we simultaneously calibrated $\sigma_z$ and $\bp$, finding not significant changes in $\bp$ and $\sigma_z=0.003^{+0.002}_{-0.002}$, which is in line with the fact that most of the cluster redshifts are spectroscopic. In addition, from the analysis of the Planck union catalogue of clusters, we showed that our result does not depend on the adopted cluster detection algorithm. We also found that a redshift evolution of $\bp$ is not necessary to describe our clustering measurements.\\
\indent By adopting priors on $\bp$ from external data sets, we found results on $\Omega_{\rm m}$ that are fully in agreement and competitive, in terms of uncertainties, with those derived from cluster counts by \citet{PlanckCluserCounts15}, while $\sigma_8$ is not constrained. By assuming cluster clustering and counts to be statistically independent, we found that their combination provides a reduction of up to $\sim30\%$ in the $\Omega_{\rm m}$ uncertainty derived from counts. Future stage-4 CMB experiments \citep{Abazajian16} will detect about $10^5$ galaxy clusters through SZ effect, significantly enhancing the cluster statistical analyses. This will improve the calibration of the hydrostatic mass bias from cluster clustering, and will possibly shed light on the degeneracy between $\sigma_8$ and mass bias. In fact, such degeneracy cannot be investigated with current data since $\sigma_8$ is not constrained, as we detailed in Section \ref{sec:results:cosm}. As a consequence, along with cluster abundance, cluster clustering will play a crucial role in the understanding of the current cosmological tensions between early and late Universe observations.

\section*{Acknowledgements}
We acknowledge support from the grants PRIN-MIUR 2017 WSCC32 and ASI n.2018-23-HH.0. GC thanks the support from INAF theory Grant 2022: Illuminating Dark Matter using Weak Lensing by Cluster Satellites, PI: Carlo Giocoli. MS acknowledges financial contributions from contract ASI-INAF n.2017-14-H.0 and contract INAF mainstream project 1.05.01.86.10.

\bibliography{bib}

\end{document}